\documentclass[aip,graphicx]{revtex4-1}

\usepackage{graphicx}% Include figure files
\usepackage{dcolumn}% Align table columns on decimal point
\usepackage{bm}% bold math
\usepackage[utf8]{inputenc}
\usepackage[T1]{fontenc}
\usepackage{mathptmx}
\usepackage{etoolbox}
\usepackage{color}
\usepackage{amsmath}
\draft 
\begin{document}

%\preprint{}

\title{Disorder-induced Dark States Line Shape in Pump-Probe Spectroscopy of Polaritons} %Title of paper



\author{Luca Nils Philipp}
\affiliation{Institut für Physikalische und Theoretische Chemie, Universität Würzburg, Emil-Fischer Straße 42, 97074 Würzburg, Germany}%
\author{Julian Lüttig}
\affiliation{Department of Physics, University of Ottawa, Ontario, Canada K1N 6N5}%
\author{Roland Mitri\'c}
\email{roland.mitric@uni-wuerzburg.de}
\affiliation{Institut für Physikalische und Theoretische Chemie, Universität Würzburg, Emil-Fischer Straße 42, 97074 Würzburg, Germany}


\date{\today}

\begin{abstract}
The formation of hybrid light--matter states called polaritons provides a route to shape the photophysics and photochemistry of molecules. Accordingly, the dynamics of polaritons following photoexcitation is extensively studied. In particular, the role of the dark state manifold in such processes remains unclear. Here, we investigate the line shape of pump-probe spectra of polaritons emerging at the dark states energy under the influence of disorder. Previously, we already investigated the pump-probe line shapes of polaritons in a disorder-free model and identified distinct signatures of relaxation into dark states, thus providing an indirect probe of this relaxation. Since the transition dipole moment of dark states vanishes in a disorder-free model, they cannot be directly probed. However, dark states acquire small transition dipole moments as soon as disorder is explicitly included in the model enabling them to be directly probed. In this work, we demonstrate that the inclusion of the relaxation to dark states and disorder leads to the evolution of the spectral shape at the DS energy from a derivative-like into an absorptive line shape. Furthermore, we investigate the dependence of the line shape on the disorder strength and its asymptotic scaling for a large number of coupled molecules. Our results demonstrate that probing dark states can help to single out the polaritonic response from the total signal.
\end{abstract}


\maketitle %\maketitle must follow title, authors, abstract and \pacs

\section{Introduction}

In recent years, the mixing of molecular excitations with light modes has become a powerful strategy to alter various molecular processes. The hybrid quasi-particles that are formed in this way are called polaritons and have by nature both properties of light and matter. Examples of processes that can be modulated by light-matter hybridization include but are not limited to energy transfer in organic semiconductors,\cite{schachenmayer_cavity-enhanced_2015, feist_extraordinary_2015, fowler-wright_mapping_2026, zhong_energy_2017, du_theory_2018, coles_polariton-mediated_2014, reitz_energy_2018, khazanov_embrace_2023} chemical reaction rates,\cite{lather_cavity_2019, thomas_ground-state_2016, hutchison_modifying_2012, schwartz_reversible_2011, sau_modifying_2021, thomas_tilting_2019, ahn_modification_2023} delocalization,\cite{quenzel_plasmon-enhanced_2022} and relaxation pathways.\cite{finkelstein-shapiro_understanding_2021, schwartz_polariton_2013, virgili_ultrafast_2011, chen_tracking_2025, groenhof_tracking_2019}

Different experimental approaches have been developed that allow control over the degree of mixing between photonic and molecular states. One can, for example, place molecules inside an optical microcavity, thereby coupling the cavity photon modes to the molecular excited states.\cite{coles_polariton-mediated_2014, delpo_polariton_2021, zhong_non-radiative_2016} Recent examples include the coupling of the excited states of the light-harvesting 2 complex to a microcavity, which alters the energetic structure and changes energy-transfer pathways, \cite{wu_optical_2022, wu_efficient_2025} as well as the formation of polariton states that enable delocalization between spatially separated donor and acceptor molecules.\cite{russo_direct_2024} Another strategy of polariton formation employs the coupling between surface-plasmon polaritons and the molecular excited states. The resulting quasi-particles are referred to as plexcitons. The degree of mixing between the two components allows one to fine tune the energetic structure and thus control the specific process. Such control is possible by choosing an appropriate experimental geometry such as the Kretschmann geometry that allows one to control the degree of mixing by the geometric angle between the light beam and the sample surface.\cite{rodel_role_2022} Although the degree of mixing between photonic and molecular states can be tuned precisely through the experimental geometry, controlling the molecular contribution is far more challenging. Relevant parameters include the number of molecules participating in the coupling process and, in extended molecular systems such as polymers and aggregates, the geometric structure of the participating chromophores.\cite{lambert_energy_2015, turkin_fluorescence_2021} In addition, the energetic structure of molecular systems is also influenced by disorder too. The latter can induce variations in molecular properties across an ensemble, arising from differences in local environments, conformations, and intermolecular interactions. These variations result in distributions of site energies, transition dipole moments, and coupling strengths, which in turn affect optical spectra, excitation dynamics, and transport.

As the number of experimental studies increases, substantial theoretical effort has been devoted to understanding polariton formation and dynamics.\cite{ribeiro_polariton_2018, sanchez-barquilla_theoretical_2022} However, several effects that are difficult to describe theoretically must be taken into account to bridge the gap between experiment and theory. Theoretical approaches often consider idealized systems to reduce complexity while retaining the essential physics.\cite{reitz_nonlinear_2025} Theoretical investigations of the influence of polariton formation on molecular photophysics and photochemistry are complicated by the fact that the experimental realizations of polariton formation almost always require the coupling of a large number of molecules to the electromagnetic field mode. Especially when trying to incorporate disorder within the theoretical model, one often cannot avoid numerical diagonalization of the Hamiltonian matrix to obtain all eigenstates and the corresponding energies. However, due to the exponential scaling of the Hilbert space dimension with the number of molecules, the diagonalization cannot be carried out for realistic system sizes. Even the diagonalization of only a small subspace of the full Hilbert space can become computationally infeasible even for relatively small systems. Therefore, if resorting to simulations incorporating only a small number of coupled molecules, it is essential to verify that the conclusions remain robust as the system size is gradually increased.

When polaritons are formed, a manifold of purely molecular states also emerges simultaneously. Since these states cannot get excited by external light sources in an ideal system due to their vanishing photonic character, they are usually called dark states (DS). However, when variations in the molecular site energies are taken into account, the DS acquire small photonic weights. Therefore, they can get excited by external light sources and thus they are sometimes also called gray states.\cite{botzung_dark_2020,parolin_molecularly_2024} Since the number of DS greatly exceeds the number of polaritonic states, relaxation from the polaritonic states to DS is likely \cite{scholes_entropy_2020} and provides an explanation for the observation of molecular-like properties in hybrid light--matter systems.\cite{schwartz_polariton_2013, gonzalez-ballestero_uncoupled_2016, botzung_dark_2020, georgiou_generation_2018, xiang_state-selective_2019}

In our recent work, we showed how energy transfer to dark states results in specific signatures in transient absorption spectroscopy.\cite{philipp_line_2025} We combined perturbation theory describing the light--matter interaction with a Tavis--Cummings (TC) model allowing us to identify the typical line shape changes connected to energy transfer to dark states. In the presented work, we extend our description to include molecular disorder. We show that this small change in the parameter space has significant impact on the line shape of the transient absorption spectra.

\section{Theoretical Methods}

\subsection{Tavis--Cummings model}
\begin{figure*}
  \includegraphics[]{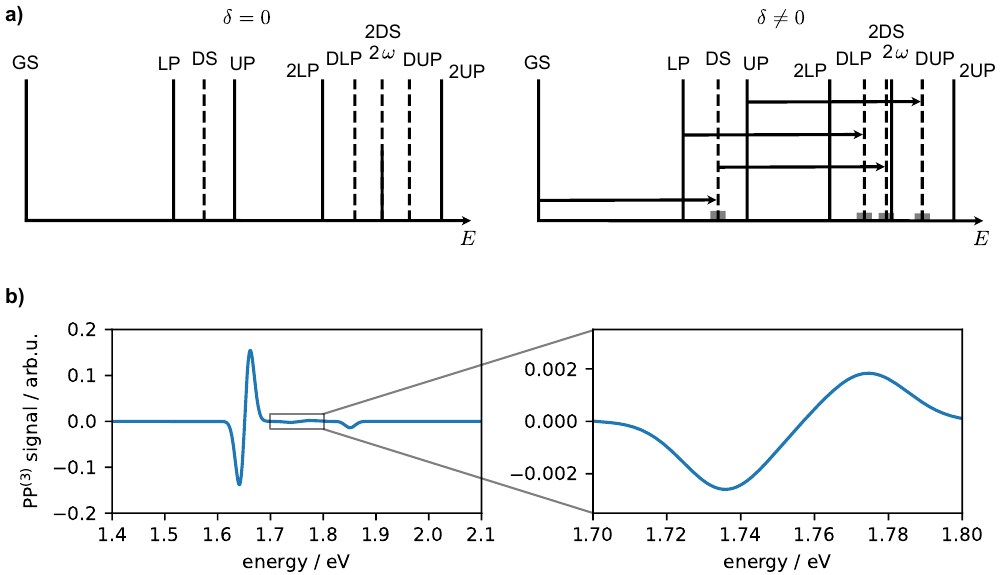}
  \caption{\label{figure1}a) Energy-level structure of the one- and two particle states of the TC model with and without disorder. States with purely polaritonic character are indicated with solid horizontal lines, while states with partial DS character are indicated by dotted horizontal lines. Only DS-type transitions, which become allowed by the introduction of on-site disorder, are indicated with arrows. The grey bars indicate the corresponding transition strength. b) Pump--probe spectrum of polaritons pumped at the LP energy and probed over the whole spectral range. The inset shows the signal around the DSs energy.}
\end{figure*}
In order to keep the physics simplest as possible we consider the TC model consisting of a single electromagnetic field mode and $N$ non-interacting two-level molecules.\cite{tavis_exact_1968, tavis_approximate_1969} We extend the model by including on-site disorder represented by random variations in the excitation energies of the individual molecules. The corresponding Hamiltonian is given by
\begin{equation}
    \label{eq:TC model}
    H_\mathrm{TC} = \hbar\sum_{i=1}^N (\omega_m+\delta_i)\sigma_i^+\sigma_i^- + \hbar\omega_ca^\dag a + \hbar g \sum_{i=1}^N\left(a^\dag\sigma_i^- + \sigma_i^+a\right),
\end{equation}
where $g$ is the light--matter coupling constant, $a^\dagger$ ($a$) is the creation (annihilation) operator of the electromagnetic field mode of energy $\hbar\omega_c$, and $\sigma_i^+$ ($\sigma_i^-$) is the creation (annihilation) operator of the \textit{i}-th molecule with excitation energy $\hbar\left(\omega_m+\delta_i\right)$. To include onsite molecular disorder we draw $\delta_i$ for every molecule from a uniform distribution on $[-\frac{\Delta}{2\hbar}, \frac{\Delta}{2\hbar}]$ with the disorder width $\Delta$ as in previous theoretical works.\cite{botzung_dark_2020, philipp_excitonexciton_2026} We expect that the choice of the distribution does not qualitatively change the results.\cite{philipp_excitonexciton_2026} Since the TC Hamiltonian commutes with the generalized number operator $n = a^\dag a + \sum_{i=1}^N\sigma_i^+\sigma_i^-$, the corresponding Hilbert space can be separated into subspaces, where every subspace is characterized by a particular excitation number. Interactions of the system with an external electromagnetic field lead to transitions between sectors with different particle numbers. If the external fields are weak, each interaction can only change the excitation number by $\pm1$. Thus, the simulation of pump--probe (PP) spectra, where the system interacts twice with the pump pulse and once with the probe pulse, necessitates the knowledge of the single- and two-particle eigenstates of the TC Hamiltonian in the simplest approximation.

If there is no disorder within the molecular ensemble and if the cavity is in resonance with the molecular transition, analytical expressions of the one- and two-particle eigenstates of the TC model and their energies are well known.\cite{delpo_polariton_2020} The one-particle states consist of the upper polariton (UP), the lower polariton (LP), and $N-1$ dark states. While DS are purely molecular states in an ideal system and have energies corresponding to the purely molecular transition energy, LP and UP have hybrid light--matter character and their energies split up symmetrically around the resonance energy. However, when on-site disorder of molecular energies is included in the TC model, DS acquire small photonic character. 

The eigenstates within the two-particle subspace can be characterized by two simultaneous excitations of the one-particle eigenstates. Thus, they can be classified as polariton--polariton, polariton--dark, and dark--dark states. There are three different non-degenerate polariton--polariton states, the second lower polariton (2LP), the second upper polariton (2UP), and the $2\omega$ state. While the 2LP (2UP) can be characterized as two simultaneous excitations of the LP (UP), the $2\omega$ state is a simultaneous excitation of LP and UP. The polariton--dark states consist of two classes of $N-1$-fold degenerate states, the dark lower polaritons (DLP), and dark upper polaritons (DUP). Just like the ground to dark state transitions, transitions from LP (UP) to one of the DLPs (DUPs) are only forbidden as long as no on-site disorder is included in the Hamiltonian. Dark-dark states only consist of a manifold of purely molecular states at the energy of twice the molecular resonance energy, so-called 2DS. The inclusion of on-site disorder also leads to partially allowed transitions from DSs to the 2DSs and LP to DLPs. The energy level structure of the one- and two-particle eigenstates of the TC model is sketched in Fig. \ref{figure1}a. For a more detailed analysis of the eigenstates of the TC model and the transitions between those, we refer to our prior publication.\cite{philipp_line_2025}

\subsection{Simulation of transient spectra}
Based on the eigenstate structure of the TC model, corresponding PP spectra can be calculated within the response function formalism \cite{mukamel_principles_1995, hamm_concepts_2011}. Here, we will only briefly outline the general idea for calculating PP spectra of polaritons and refer to our prior publications for a more detailed treatment \cite{buttner_probing_2025,philipp_line_2025}.

The total system Hamiltonian including the classically treated external electromagnetic field $E(t)$ of the driving laser pulses is given by
\begin{equation}
    H = H_{\mathrm{TC}} - \mu E(t).
\end{equation}
Here, we will assume that only the photonic degrees of freedom are driven by the external electromagnetic field, in which case the transition moment is given by $\mu = \lambda\left(a+a^\dagger\right)$, where $\lambda$ is a constant connected to the coupling strength between the electromagnetic field mode within the cavity and the exciting external electromagnetic field.\cite{tame_single-photon_2008} The interactions of the system with the external electromagnetic fields of the laser pulses drives the system in a non-equilibrium state. Subsequently, this state emits the macroscopic polarization, which is connected to the system's density matrix by
\begin{equation}
    P(t) = \mathrm{Tr}\left(\mu\rho(t)\right).
\end{equation}
Since the external laser field is regarded as weak, the system's density matrix can be perturbatively expanded in the interaction with the external field,
\begin{equation}
    \rho(t) = \sum_{n=0}^\infty \rho^{(n)}(t).
\end{equation}
The perturbative expansion of the density matrix induces the same kind of expansion for the polarization. In transiently detected PP spectroscopy, the leading nonlinear order in the perturbative expansion of the polarization is the third order. Within the semi-impulsive limit, it is assumed that the temporal width of the laser pulses is small compared to any relaxation process in the system while retaining the carrier frequency and wavevector of the fields. By invoking this approximation, the third-order polarization becomes proportional to the third-order response function $S^{(3)}(T, t)$,
\begin{equation}
    P^{(3)}(T, t) \propto S^{(3)}(T, t),
\end{equation}
with the delay time $T$ between pump and probe pulse and the detection time $t$ after which the polarization is detected.

The third-order response function is given by 
\begin{equation}
    S^{(3)} (T, t)=\left(-\frac{i}{\hbar}\right)^3 \mathrm{Tr}\left(\mu U(t)[\mu,U(T)[\mu,[\mu, \rho_{\mathrm{eq}}]]]\right),
\end{equation}
where $\rho_{\mathrm{eq}}$ is the density matrix in thermal equilibrium and $U$ is the propagator governing the time evolution of the density matrix. The nested commutators in the expression of the response function results in various terms contributing to the response of the system. The appearance of $n$ commutators in the response function leads to $2^n$ contributing terms. However, among these terms only pairs being complex conjugates of each appear, which already halves the number of effective terms. Furthermore, the number of contributing terms can be reduced by invoking different approximations, i.e., the rotating wave approximation and time ordering, and applying experimental techniques such as phase matching.\cite{hamm_concepts_2011} These terms are commonly referred to as Liouville space pathways and can be graphically represented by double-sided Feynman diagrams.\cite{mukamel_principles_1995} In double-sided Feynman diagrams time is running from the bottom to the top. Starting from the ground state, the system first interacts simultaneously twice with the pump pulse indicated by solid arrows. Of these two interactions with the pump pulse, one occurs with $-k_{\mathrm{pu}}$ and the other one with $+k_{\mathrm{pu}}$. During the delay time $T$ the system might decay, which is indicated by a dotted horizontal line. Subsequently, a coherence is created by the interaction with the probe pulse with wavevector $+k_{\mathrm{pr}}$. The coherence oscillates during the detection time $t$ until the polarization is emitted indicated by a dotted arrow. Since every interaction from the right carries a minus sign due to the commutator, each Liouville space pathway carries the sign $(-1)^m$, where $m$ is the number of interactions from the right.

In the case of coherently detected PP spectroscopy, there are only three different kinds of Liouville space pathways, i.e., ground state bleach (GSB), stimulated emission (SE), and excited state absorption (ESA). In GSB pathways, the system remains in the ground state after the first two interactions with the pump pulse. Contrasting that, the system is pumped into an excited state for SE and ESA pathways. While the system is brought back to the ground state by the probe pulse in SE pathways, in ESA pathways the system is further excited by the probe pulse from a one- into a two-particle eigenstate. 

By summing up all contributing Liouville space pathways the total response function is obtained, which is subsequently Fourier transformed to obtain a quantity, whose imaginary part is proportional to the third-order PP spectrum $\mathrm{PP}^{(3)}(\omega)$ \cite{mukamel_principles_1995, hamm_concepts_2011},
\begin{equation}
    \mathrm{PP}^{(3)}(\omega) \propto -2\mathrm{Im}[P^{(3)}(\omega)].
\end{equation}
As an example of how a Liouville space pathway finally contributes to the PP spectrum, the expression contributing by the $\mathrm{ESA_{2DS}}$ pathway is given by
\begin{equation}
    \mathrm{PP^{(3)}_{ESA_{2DS}}}(\omega) \propto \mu_\mathrm{LP}^2 U_\mathrm{LP,DS}(T)\mu_\mathrm{DS\rightarrow2DS}^2 \delta(\omega_\mathrm{2DS}-\omega_\mathrm{DS}).
\end{equation}
Since we neglect decay of the coherence during the detection time, the spectral line shape has delta form. Here, we assume that inhomogeneous broadening is dominant. Therefore, we explicitly include onsite disorder in the TC Hamiltonian (Eq. (\ref{eq:TC model}), replace the delta function by a Gaussian line shape with a narrow homogeneous width of $\sigma = 0.01\ \mathrm{eV}$, and average over 1000 disorder realizations.

We consider here only the dependence of population states during the delay time $T$, while neglecting any coherences which often dephase quickly. Therefore, it is convenient to use rate equations to calculate the matrix elements of the propagator. Here, we use the following system of differential equations to describe the time dependence of the populations $P_i = \rho_{ii}$,
\begin{gather}
    \frac{\mathrm{d}P_\mathrm{LP}}{\mathrm{d}T} = -(k_\mathrm{LP} + (N-1)k_\mathrm{DS})P_\mathrm{LP}\\
    \frac{\mathrm{d}P_{\mathrm{DS}_i}}{\mathrm{d}T} = k_\mathrm{DS}P_\mathrm{LP}\\
    \frac{\mathrm{d}P_\mathrm{GS}}{\mathrm{d}T} = k_\mathrm{GS}P_\mathrm{LP}.
\end{gather}
This system of differential equations describes a parallel decay of the LP with rate $k_\mathrm{DS}$ into each of the $N-1$ dark states and into the ground state with rate $k_\mathrm{GS}$. The analytical solution is easily obtained and if only the LP is initially populated it is given by
\begin{gather}
    P_\mathrm{LP}(T) = e^{-(k_\mathrm{LP} + (N-1)k_\mathrm{DS})t}\\
    P_{\mathrm{DS}_i}(T) = \frac{k_\mathrm{DS}}{(N-1)k_\mathrm{DS} + k_\mathrm{LP}}\left(1-e^{-(k_\mathrm{LP} + (N-1)k_\mathrm{DS})t}\right)\\
    \mathrm{d}P_\mathrm{GS}(T) = \frac{k_\mathrm{LP}}{(N-1)k_\mathrm{DS} + k_\mathrm{LP}}\left(1-e^{-(k_\mathrm{LP} + (N-1)k_\mathrm{DS})t}\right).
\end{gather}
The overall decay rate of the LP is given by $k_\mathrm{LP} = k_\mathrm{GS} + (N-1)k_\mathrm{DS}$. The system of differential equations can also be cast into matrix form,
\begin{equation}\label{propagator}
    \frac{\mathrm{d}}{\mathrm{d}T}P = KP,
\end{equation}
where $P$ is the vector containing all populations and $K$ is the matrix with all rate constants. Then, the matrix elements of the propagator are given by $U_{if}(T)=(\mathrm{exp}\left(KT\right))_{if}$ and the matrix elements correspond to the probability that given the system is initialized in the initial state $i$ it is in the final state $f$ after time $T$.

\section{Results and Discussion}
\subsection{Origin of the disorder-induced dark state line shape}
\begin{figure*}
  \includegraphics[]{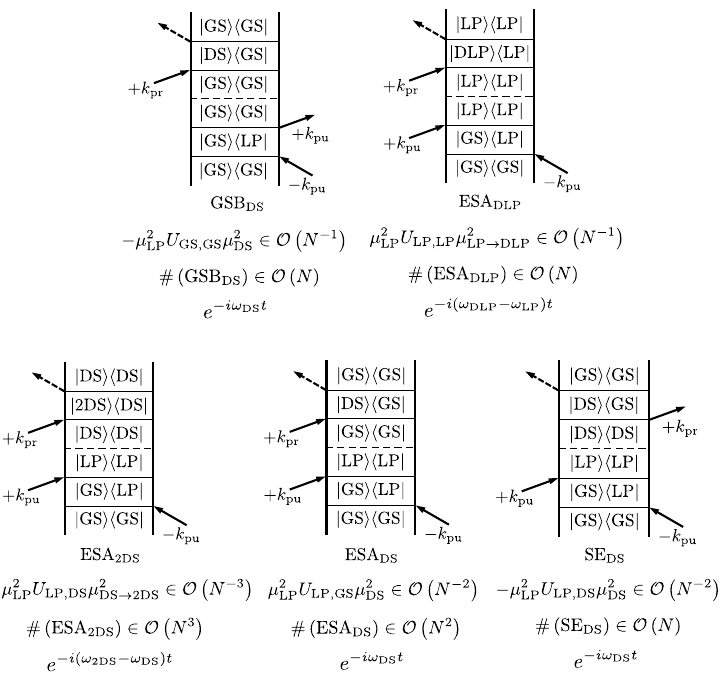}
  \caption{\label{figure2} Double-sided Feynman diagrams representing the Liouville space pathways, which describe the nonlinear response of polaritons pumped at the LP energy and emit the signal approximately at the energy of the DSs. Below every diagram the amplitude of the diagram together with its asymptotic scaling for large number of molecules, the asymptotic scaling of the number of diagrams, and the time-dependent phase factor accumulated during the time detection time are shown.}
\end{figure*}
By the introduction of on-site disorder within the TC model otherwise forbidden transitions at the DSs energy acquire small transition amplitudes, which leads to a small signal at the dark states energy (Fig. \ref{figure1}b). Consequently, a series of Liouville space pathways, which contain DS transitions, become important to properly describe the response of the polaritonic system. Here, we assume that the pump pulse is resonant with the LP, while the probe pulse is centered around the energy of the DSs such that only transitions around the DSs energy are probed. The resulting pathways are depicted in Fig. \ref{figure2}. While the $\mathrm{GSB_{DS}}$ Liouville space pathway always contributes to the systems response regardless of whether the LP relaxes fast into the GS and DSs or not, the $\mathrm{ESA_{DLP}}$ pathway only contributes if the system is not fully relaxed. In contrast, the remaining three pathways, $\mathrm{ESA_{2DS}}$, $\mathrm{ESA_{DS}}$, and $\mathrm{SE_{DS}}$, only contribute when the LP does relax into the GS and DSs.
\begin{figure*}
  \includegraphics[]{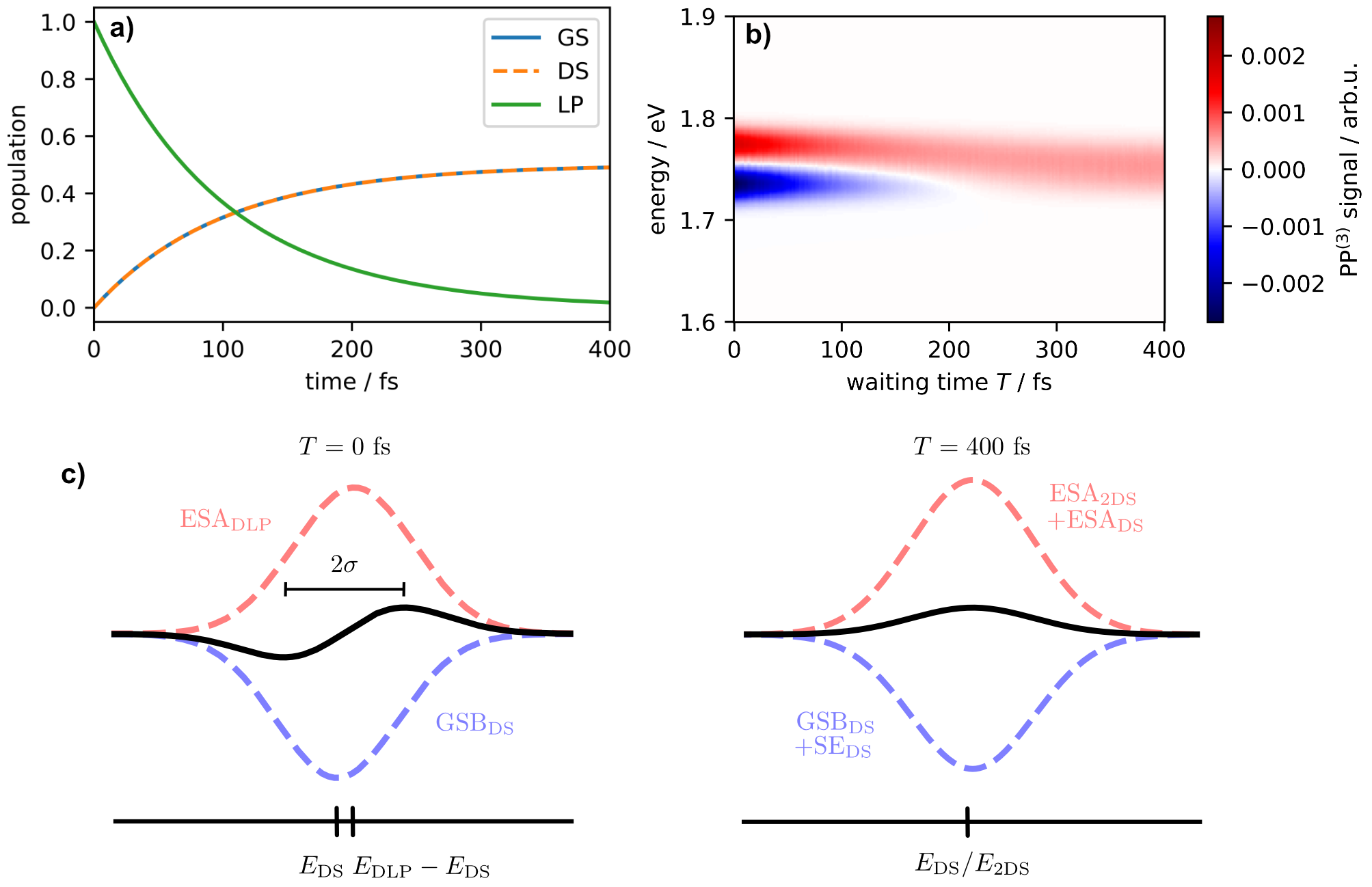}
  \caption{\label{figure3} a) Population dynamics of the kinetic polariton relaxation model. b) Time-dependence of the PP spectrum of polaritons pumped at the LP energy and probed at the DSs energy. During the delay time the initially pumped LP state relaxes with equal share into the manifold of DSs and into the GS. c) Schematic representation of the formation of the derivative-like line shape at early delay times (left panel) and schematic representation of the formation of the trivial line shape at late delay times (right panel).}
\end{figure*}
To obtain the third-order PP spectra, the propagator is calculated for each delay time $T$ with the parameters $\frac{1}{k_\mathrm{LP}}=100\ \mathrm{fs}$, $\frac{1}{k_\mathrm{GS}}=50\ \mathrm{fs}$, and $\frac{1}{k_\mathrm{DS}}=\frac{50}{N-1}\ \mathrm{fs}$. For these parameters, the LP relaxes with a time constant of $100\ \mathrm{fs}$ into the ground state and the DSs, where 50\% of the LP population relaxes back to the ground state and 50\% relaxes into the manifold of DSs (Fig. \ref{figure3}a). Furthermore, the eigenstates of the disordered TC Hamiltonian and the transition amplitudes between the states are calculated by diagonalizing the Hamiltonian for the parameters $\hbar\omega_m=\hbar\omega_c=1.75\ \mathrm{eV}$, $\hbar g\sqrt{N}=0.1\ \mathrm{eV}$, and $\Delta=0.05\ \mathrm{eV}$. Based on this, the third-order PP spectra around the DSs energy are obtained by summing up the contributions of all relevant Liouville space pathways (Fig. \ref{figure2}). In Fig. \ref{figure3}b the resulting PP spectra are shown for $N=10$ as a function of the delay time $T$. While the PP spectrum has a derivative-like line shape at early times, it transiently evolves into a single positive signal. In PP spectroscopy, derivative-like line shapes are characteristic for polaritons\cite{philipp_line_2025, delpo_polariton_2020,schwartz_polariton_2013,renken_untargeted_2021}. Experimentally, this type of line shape is frequently observed at the energies of LP and UP. In our previous work, we found that the derivative-like line shape at the energies of LP and UP can be explained by a partial cancellation of positive ESA pathways and negative GSB and SE pathways due to a small energetic difference of the involved transition energies. Similarly, the derivative line shape around the energy of the DSs can be explained: When no DS relaxation has yet occurred, only the positive $\mathrm{ESA_{DLP}}$ and negative $\mathrm{GSB_{DS}}$ pathways contribute to the nonlinear response of the system. While both type of pathways emit the polarization approximately at the energy of the DSs, the energy of the DS to DLP transitions is on average slightly larger than that of the GS to DS transitions. Thus, the contributions of both type of pathways cancel each other strongly leading to the observed derivative-like line shape (Fig. \ref{figure3}c).
%Some experimental studies argue that the derivative-like spectral features at LP and UP energies arise due to an effective reduction in the number of molecules coupled to the electromagnetic field mode induced by the pump pulse. While this may serve as an phenomenological description of the PP spectra, it was already argued that it should be better regarded as phase space filling, i.e., a shift of the one- to two-particle state transition energies compared to the transition energies of the ground to one-particle state transitions due to the fact that the two-particle eigenstates are no simple tensor product states of the one-particle states. Since the by phase-space filling arising energy difference of the two transition types is small compared to the width of the line shape, the distance of the extremal points of the derivative-like line shape is given by $2\sigma$, where $\sigma$ is the standard deviation of the gaussian line shape function. In particular, the distance of the extremal points is not connected to any kind of effective number of molecules coupled to the cavity mode. This generally applies to cases, where two different spectral contributions with gaussian line shape with opposite sign strongly overlap due to the fact that their energy difference is smaller than the width of the individual contributions.

Compared to the derivative-like line shape at early times, at late delay times the line shape of the third-order PP spectra only consists of a single positive signal (Fig. \ref{figure3}c). Here, no derivative-like line shape is observed, since the amplitude of the positive $\mathrm{ESA}_\mathrm{DLP}$ pathway has completely diminished at late times due to the relaxation of the LP. Furthermore, the remaining pathways, $\mathrm{ESA}_\mathrm{DS}$, $\mathrm{ESA}_\mathrm{2DS}$, $\mathrm{GSB}_\mathrm{DS}$, and $\mathrm{SE}_\mathrm{DS}$, all emit the polarization at the energy of DS and 2DS, which are equal on average. %Since the remaining Liouville space pathways on average all emit the polarization at the same energy, no derivative-like spectral feature is formed, and the height of the single positive signal is simply determined by the sum of the amplitude of the contributing pathways.
\begin{figure*}
  \includegraphics[]{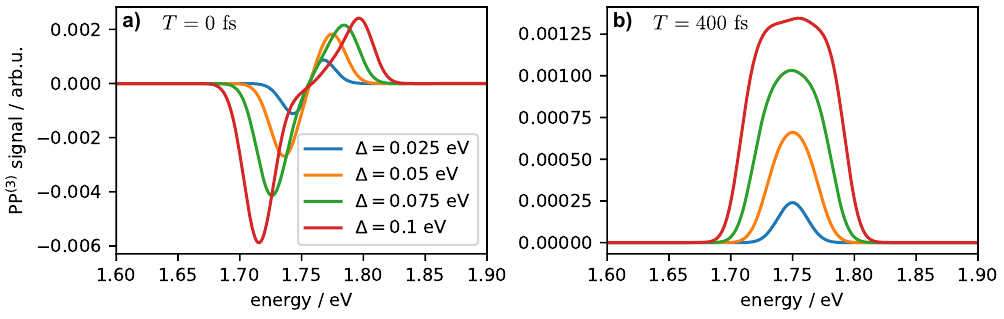}
  \caption{\label{figure5} Influence of different disorder widths on the line shape in the PP spectrum at the DSs energy for a) the derivative-like line shape at early delay times ($T=0\ \mathrm{fs}$) and b) the trivial line shape at late delay times ($T=400\ \mathrm{fs}$).}
\end{figure*}
Shown in Fig. \ref{figure5} are the PP line shapes at early and late times for various disorder widths. Generally, the signal strength rises with increasing disorder width for the line shape at early and late delay times (Fig. \ref{figure5}a and \ref{figure5}b). This increase in the signal strength occurs due to larger transition moments of the included DS-type transitions. Furthermore, the width of the signals rises due to the increased width of the DS manifold for large disorder. In particular, this can be observed by an increasing distance of the extrema of the derivative-like line shape at early times for larger disorder strengths. In a simplified model, where only two Gaussians with opposite signs, centered around slightly different energies contribute to the line shape, one can show that the difference of the two extrema of the derivative-like line shape is given by $2\sigma$, where $\sigma$ is the standard deviation of the Gaussian line shape function. Thus, one might estimate the width of the DS manifold by the difference of the two extrema of the derivative line shape. However, this estimate only works if the homogeneous width of a single signal is smaller than the inhomogeneous width induced by the different energies of the DSs. From Fig. \ref{figure5}b it is also evident that if the inhomogeneous broadening due to DSs transitions with different energies is larger than the homogeneous width of a single transition the line shape deviates from a Gaussian shape to a broader shape, which falls of steeper at the edges.

%Some experimental studies also argue that the derivative-like spectral features at LP and UP energies arise due to an effective reduction in the number of molecules coupled to the electromagnetic field mode induced by the pump pulse. While this may serve as an phenomenological description of the PP spectra, it was already argued that it should be better regarded as phase-space filling, i.e., a shift of the one- to two-particle state transition energies compared to the transition energies of the ground to one-particle state transitions due to the fact that the two-particle eigenstates are no simple tensor product states of the one-particle states. Since the by phase-space filling arising energy difference of the two transition types is small compared to the width of the line shape and as we showed actually the width of the contributions determines the distance in this case, the difference of the extrema should not be utilized to estimate an effective number of molecules coupled to the field mode.

\subsection{Asymptotic scaling of the line shape}

In our previous work on the PP line shapes of polaritons, we already argued that the amplitude of the PP signal should decrease when the number of coupled molecules is increased. This can be rationalized by considering that the ensemble of two-level systems in the TC Hamiltonian can be mapped onto a single pair of bosonic creation and annihilation operators by the Holstein-Primakov transformation. In this way, one shows that the TC model is asymptotically equivalent to a system of two coupled harmonic oscillators. Since in general, the third-order response of a system of coupled harmonic oscillators vanishes, the amplitude of the PP line shape should decrease with increasing number of molecules. This includes the amplitude of the signal at the energy of the DSs. The question is now whether the amplitude of the PP line shape at the energy of the DS has a different asymptotic large $N$ scaling than the amplitude of the PP line shape at the energy of LP and UP. 

\begin{figure*}
  \includegraphics[]{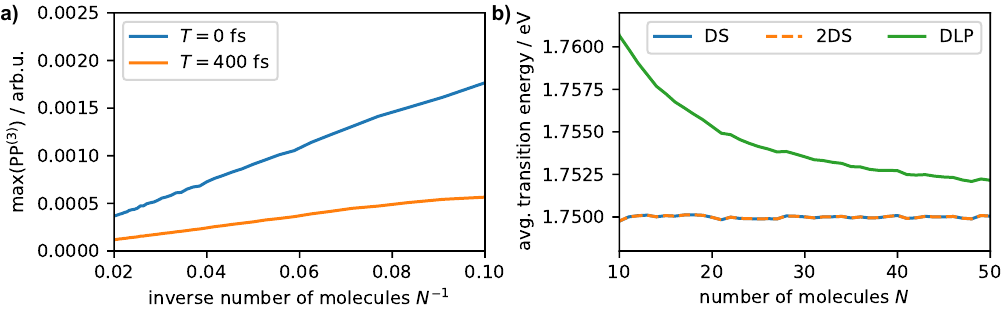}
  \caption{\label{figure4}a) Maximum of the PP spectrum at the DSs energy depending on the inverse number of molecules for the derivative-like line shape at early times ($T=0\ \mathrm{fs}$) and the trivial line shape at late times ($T=400\ \mathrm{fs}$). b) Dependence of the average transition energy of the DS-type transitions, i.e., GS to DS, DS to 2DS, and LP to DLP transitions, one the number of molecules. In the legend, the different transition types are indicated by their final state.}
\end{figure*}

In Fig. \ref{figure4}a, we show the dependence of the maximum of the PP line shape at the DS energy on the inverse number of coupled molecules for the derivative-like line shape at early delay times and the trivial line shape at late delay times. While the maximum of the derivative-like line shape is generally larger than that of the trivial line shape, it is evident that the maximum of both PP line shapes depends linearly on the inverse number of coupled molecules, i.e., the magnitude of both DS line shapes scale with an order of $\mathcal{O}\left(N^{-1}\right)$. Slight deviations from the linear behavior are observed if the number of molecules is small. For the PP signal at the energy of LP and UP, which we investigated in our recent publication \cite{philipp_line_2025}, one also finds that the maximum of this signal scales with an order of $\mathcal{O}\left(N^{-1}\right)$. These scalings can be rationalized by considering a simplified model, where there are two contributions to the line shape at energies $E$ and $E+\delta$ with opposite signs, amplitudes $A$ and $B$ with $A>B>0$, and Gaussian line shape with width $\sigma \gg \delta$. In this case, the line shape function around the energy $E$ (LP, UP, or DS) is approximately given by (see appendix A)
\begin{equation}
    \mathrm{PP}^{(3)}\left(\omega-\frac{E}{\hbar}\right) \approx \mathrm{exp}\left(-\frac{\omega^2}{2\sigma^2}\right) \left((A-B) + \frac{B\delta}{\sigma^2}\omega\right).
\end{equation}
This line shape function contains two contributions, the trivial line shape and the derivative-like line shape. The trivial line shape is a simple Gaussian with a factor given by the difference of the amplitude of both contributions $A-B$. The derivative-like line shape on the other hand, is a Gaussian multiplied with a linear function $\frac{B\delta}{\sigma^2}\omega$, which resembles the characteristic derivative-like line shape of polaritons. We found that since the derivative-like line shape is often the dominant contribution, the trivial part of the line shape only dominates if the derivative-like line shape is absent, i.e., $\delta=0$ or $B=0$. For the PP spectrum line shape at the DSs energy, the derivative-like line shape obviously dominates at early times, while for late times, the amplitude of the derivative-like contribution is diminished leading to a trivial line shape.

The form of the approximated line shape function reveals that the asymptotic scaling of the trivial line shape is determined by the scaling of the amplitude difference $A-B$. The asymptotic scaling of the derivative-like line shape on the other hand is determined by the pathway amplitude $B$ and the energy offset $\delta$. Here, the energy offset comes into play, since it determines how much the two signals of opposite signs cancel each other out. Using the scaling of these parameters, we can show why the signal at the energy of the LP scales as $\mathcal{O}\left(N^{-1}\right)$. There are three different Liouville space pathways contributing to this signal. Hereby, stimulated emission of the LP ($\mathrm{SE_{LP}}$) and ground state bleach at the LP energy ($\mathrm{GSB_{LP}}$) correspond to one contribution in the model, and excited state absorption from LP to 2LP ($\mathrm{ESA_{LP\rightarrow 2LP}}$) represents the contribution with slight energy offset due to the slightly different transition energies. As we show in the appendix, the amplitude of all three pathways is asymptotically independent of the number of coupled molecules. However, the difference of the amplitudes is on the order of $\mathcal{O}\left(N^{-1}\right)$. Therefore, the trivial line shape also scales on the order of $\mathcal{O}\left(N^{-1}\right)$. Since the amplitudes do not depend on the number of molecules asymptotically, the energy offset $\delta$ is the key parameter. In the appendix, we show that the energy difference of the LP to 2LP and the GS to LP transition is on the order of $\mathcal{O}\left(N^{-1}\right)$. Thus, the magnitude of the derivative-like line shape and also the approximated line shape function overall scales with an order of $\mathcal{O}\left(N^{-1}\right)$. This can be similarly shown for any signal at the LP and UP energies when pumping either LP or UP in the resonant, disorder-free TC model.

Similarly the scaling of the line shape at the DS energy in the disordered TC model can be rationalized. However, here the scaling of the amplitude of the Liouville pathways is more difficult to determine, since there are many more pathways. There are three different components in the pathway amplitudes: the matrix elements of the propagator, the transition moments, and the number of pathways for a given DS transition. Under our assumptions, the propagator matrix element $U_\mathrm{LP,DS}(T)$ is on the order of $\mathcal{O}\left(N^{-1}\right)$, since there are $N-1$ dark states. All other propagator matrix elements, $U_\mathrm{LP,LP}$, $U_\mathrm{GS,GS}$, and $U_\mathrm{LP,GS}$, are on the order of $\mathcal{O}\left(1\right)$, since there is only a single final state. Regarding the scaling of the transition moments, it is known that $\mu^2_\mathrm{LP}$ is on the order of $\mathcal{O}\left(1\right)$, while the photonic weight of a DS, which corresponds to its squared transition moment $\mu_\mathrm{DS}^2$, is on the order of $\mathcal{O}\left(N^{-1}\right)$, i.e., the transition moment scales with the inverse number of possible final states. Similarly, we have numerically checked that $\mu_\mathrm{LP\rightarrow DLP}^2$ scales with $\mathcal{O}\left(N^{-1}\right)$ and $\mu_\mathrm{DS\rightarrow 2DS}^2$ scales with $\mathcal{O}\left(N^{-2}\right)$. The scaling of the individual pathway amplitudes and the number of individual pathways for each class are shown in Fig. \ref{figure2}. It is evident that for every pathway class, except for $\mathrm{SE_{DS}}$, the scaling of the pathway amplitude is balanced out by the number of individual pathways such that the contribution of the pathway class is on the order of $\mathcal{O}\left(1\right)$. Compared to this, the $\mathrm{SE_{DS}}$ pathways become negligible if the number of molecules is large, since the $\mathcal{O}\left(N\right)$ pathways cannot balance out the $\mathcal{O}\left(N^{-2}\right)$ scaling of the individual pathway amplitudes.
To investigate the asymptotic scaling of the energy offset between different DS transition energies one has to calculate an average transition energy for each class, since there are many transitions of a single kind. Thus, we define the average transitions energy of a given transition class by
\begin{equation}
    E_{I\rightarrow F} = \frac{\sum_{i\in I}\sum_{f\in F} |\mu_{i,f}|^2 (E_f - E_i)}{\sum_{i\in I}\sum_{f\in F} |\mu_{i,f}|^2},
\end{equation}
where $I$ and $F$ are index sets containing all initial and final states of a given transition class. The resulting dependence of the transition energies on the number of molecules is shown in Fig. \ref{figure4}b. While the averaged DS and 2DS transition energies are equal for arbitrary number of molecules, the average DLP transition energy is always slightly blue shifted. However, the average DLP transition energy approaches the average transition energy of the DS and 2DS transitions with an increasing number of molecules. Closer inspection reveals that the energy difference between the DLP and DS (2DS) transitions as a function of $N$ is on the order of $\mathcal{O}\left(N^{-1}\right)$. Consequently, the derivative-like line shape also scales with $\mathcal{O}\left(N^{-1}\right)$. Finally, we conclude that the PP line shape around the DS energy shows the same asymptotic scaling as the PP line shapes at the LP and UP energies.

This conclusion assures that the PP signal around the DS energy does not fall of faster than the signal at LP and UP energies with increasing number of molecules. Thus, the DS line shape might in principal be observable for the large system sizes typically encountered in experiments. However, this does not mean that the signal at the DSs energy is on the same order of magnitude as the signals at the LP and UP energies. As evident from Fig. \ref{figure1}b the magnitude of the DSs signal might still be several orders of magnitudes smaller than that at the LP or UP energy depending on the disorder strength. Therefore, large signal to noise ratios might be necessary to observe the PP signal at the DSs energy. Nevertheless, the PP signal at the DSs energy might serve as a characteristic fingerprint feature of polaritons, since it allows us to distinguish clearly between the response of the polaritonic states (LP and UP) and the DSs by observing either a derivative-like or trivial line shape at the DSs energy.

\section{Conclusion}
In this work, we have combined the quantum mechanical Tavis--Cummings model including on-site disorder to simulate pump--probe spectra of polaritons at the DSs energy after initial excitation of the LP within the response function formalism. Although DS-type transitions are usually forbidden in the TC model, a signal at the DS energy arises if disorder is explicitly included in the Hamiltonian. We have simulated the time-dependence of the PP spectrum at the DSs energy if the initially excited LP relaxes into the DSs and the GS. Hereby, the relaxation of the LP was described by coupled rate equations. For early delay times, the PP spectrum has a derivative-like line shape which transiently evolves into a single positive peak, which we refer to as the trivial line shape. The analysis of the contributing Liouville space pathways showed that the derivative-like line shape arises due to a small energy difference between different transitions in the TC model. This energy offset is absent for the transitions included in the Liouville space pathways after relaxation of the LP such that after the relaxation only the trivial line shape appears. In addition, we have investigated the influence of disorder strength on the PP line shape at the DSs energy. By increasing the disorder, the signal at the DSs energy becomes larger due to increased DS-type transition moments and its width increases due to the energetically broader DS manifold. Since we showed that the distance of the extrema of the derivative-like line shape is directly connected to the width of the signal of the individual Liouville space pathway classes, the distance of the extrema is a direct measure of disorder in a polaritonic system if the inhomogeneous broadening by disorder is larger than the homogeneous broadening. 

Furthermore, we have constructed a simplified model, which describes the formation of the line shape and contains the derivative-like and trivial line shape as the only two contributions. With this model at hand, we were able to analyze the asymptotic scaling of the amplitude of the PP line shape with the number of molecules. In this way, we found that the amplitude of the DS line shape has the same asymptotic scaling as the line shape at the energy of LP and UP, i.e., both scale on the order of $\mathcal{O}\left(N^{-1}\right)$. Therefore, the DS line shape is in principal also observable if there is a signal at the polariton energies. However, this does not mean, that the amplitude of the signals are on the same order of magnitude, since, as we showed, the signal at the DS energy still heavily depends on the disorder strength. We expect it to be several orders of magnitude smaller than that at the energy of the polaritonic states. Thus, large signal to noise ratios might be needed to clearly observe the DS line shape.

Nevertheless, if it is observable, it carries a lot of valuable information about the system. From the line shape one can deduce whether one really observes the response of polaritons or that of the purely molecular DSs, i.e., the derivative-like line shape at the energy of the DS represents an characteristic fingerprint of polaritons.

% \begin{acknowledgments}
% We wish to acknowledge the support of the author community in using
% REV\TeX{}, offering suggestions and encouragement, testing new versions,
% \dots.
% \end{acknowledgments}

\begin{acknowledgments}
 L.N.P. acknowledges a fellowship by the \textit{Fonds der Chemischen Industrie} (FCI). J.L. acknowledges support from the HFSP fellowship program under Grant No. LT0056/2024-C.
\end{acknowledgments}

\section*{Conflicts of interest}
\vspace*{-2ex}
There are no conflicts to declare.

\section*{Data availability}
\vspace*{-2ex}
The data that support the findings of this study are available from the corresponding author upon reasonable request.

\appendix

\section{Details on Line Shape Formation}

In this section, we will investigate the formation of the PP line shapes in more detail. To this end, consider a simplified model, where there are two contributions to the line shape at energies $E$ and $E+\delta$ with opposite sign and amplitudes $A$ and $B$ with $A>B>0$. Furthermore, we assume that the line shape function of each contribution is Gaussian with width $\sigma$ and the energy offset $\delta$ is assumed to be small compared to the width $\delta\ll\sigma$. We will show that then the line shape can be decomposed into two contributions
\begin{equation}
    \mathrm{PP}^{(3)}\left(\omega-\frac{E}{\hbar}\right) = \mathrm{PP}^{(3)}_{\mathrm{triv}}\left(\omega-\frac{E}{\hbar}\right) + \mathrm{PP}^{(3)}_{\mathrm{deri}}\left(\omega-\frac{E}{\hbar}\right)
\end{equation}
with the trivial and the derivative-like line shape defined as
\begin{eqnarray}
    \mathrm{PP}^{(3)}_{\mathrm{triv}}\left(\omega-\frac{E}{\hbar}\right) &= (A-B)\ \mathrm{exp}\left(-\frac{\omega^2}{2\sigma^2}\right)\\
    \mathrm{PP}^{(3)}_{\mathrm{deri}}\left(\omega-\frac{E}{\hbar}\right) &= B\ \mathrm{exp}\left(-\frac{\omega^2}{2\sigma^2}\right) \frac{\omega\delta}{\sigma^2}.
\end{eqnarray}
The third-order PP signal for the described model around the energy $E$ is given by
\begin{equation}
    \mathrm{PP}^{(3)}\left(\omega-\frac{E}{\hbar}\right) = A\ \mathrm{exp}\left(-\frac{\omega^2}{2\sigma^2}\right) \\-B\ \mathrm{exp}\left(-\frac{(\omega-\delta)^2}{2\sigma^2}\right).
\end{equation}
Expanding the equation yields
\begin{equation}
    \mathrm{PP}^{(3)}\left(\omega-\frac{E}{\hbar}\right) = (A-B)\ \mathrm{exp}\left(-\frac{\omega^2}{2\sigma^2}\right) + B\left(\mathrm{exp}\left(-\frac{\omega^2}{2\sigma^2}\right) - \mathrm{exp}\left(-\frac{(\omega-\delta)^2}{2\sigma^2}\right)\right).
\end{equation}
While the first term is already the trivial line shape, the derivative-like line shape is obtained from the second term by approximating the exponential within the brackets linearly , which is valid for $\omega\delta\ll\sigma$, and neglecting terms on the order of $\delta^2$:
\begin{equation}
    B\ \mathrm{exp}\left(-\frac{\omega^2}{2\sigma^2}\right) \left(1- \mathrm{exp}\left(-\frac{2\omega\delta - \delta^2}{2\sigma^2}\right)\right) \approx B\ \mathrm{exp}\left(-\frac{\omega^2}{2\sigma^2}\right) \frac{\omega\delta}{\sigma^2}.
\end{equation}
From this equation, the derivative-like shape of this line shape is obvious. Furthermore, it is easily shown that this line shape function has extrema at $\omega_{1,2} \approx \pm \sigma$, when neglecting terms in the order of $\delta^2$. 

% Based on this equation it is also possible to derive a simple criterion whether the line shape is trivial or derivative like. To this end, consider the approximated line shape function including the trivial and derivative-like part

% \begin{equation}
%     \mathrm{PP}^{(3)}\left(\omega-\frac{E}{\hbar}\right) \approx \mathrm{exp}\left(-\frac{\omega^2}{2\sigma^2}\right) \left((A-B) + B\frac{\omega\delta}{\sigma^2}\right).
% \end{equation}

% Here, we define that the line shape is derivative-like if the polynomial has a root within the interval $[-\sigma, \sigma]$. This leads to the simple criterion that the line shape is derivative-like if 

% \begin{equation}
%     \left|\frac{A-B}{B}\frac{\sigma}{\delta}\right| < 1.
% \end{equation}

% When considering the asymptotic scaling of the line shape with the number of molecules in the TC model, there appear two quantities within the approximated line shape function, which depend on the number of coupled molecules, i.e., the pathway amplitudes $A$ and $B$ and the energy offset $\delta$. The collective coupling $g_c = g\sqrt{N}$ is kept constant for varying numbers of molecules, since if otherwise $g$ was kept constant, the collective coupling would diverge. 

This representation of the line shape can be used to show that the amplitude of the derivative-like spectral feature at the LP energy arising when pumping the LP asymptotically scales with $N^{-1}$. In this case, there are 3 different Liouville space pathways contributing to this signal, stimulated emission of the LP $\mathrm{SE_{LP}}$, ground state bleach at the LP energy $\mathrm{GSB_{LP}}$, and excited state absorption from LP to 2LP $\mathrm{ESA_{LP\rightarrow 2LP}}$.\cite{philipp_line_2025} Therefore, the pathway amplitudes are given by
\begin{eqnarray}
    A &= \mu^2_\mathrm{LP}\mu^2_\mathrm{LP\rightarrow 2LP} \\
    B &= 2\mu^4_\mathrm{LP} \\
    \delta &= \frac{E_\mathrm{2LP} - 2E_\mathrm{LP}}{\hbar},
\end{eqnarray}
where $E_\mathrm{LP} = \hbar\omega_m - \hbar g_c$ ($E_\mathrm{LP\rightarrow 2LP} = 2\hbar\omega_m - 2\hbar g_c\frac{\sqrt{2N-1}}{\sqrt{2N}}$) is the resonance energy of the LP (2LP), $\mu_\mathrm{LP} = \frac{1}{\sqrt{2}}$ ($\mu_\mathrm{LP\rightarrow 2LP} = \frac{1}{2} + \sqrt{\frac{N}{2(2N-1)}}$) is the transition moment of the GS to LP (LP to 2LP) transition, and $\sigma$ is the width of both contributions. In this case, the energy offset is given by
\begin{equation}
    \delta = 2\hbar g_c\left(1 - \frac{\sqrt{2N-1}}{\sqrt{2}N}\right),
\end{equation}
which is of order $\mathcal{O}\left(N^{-1}\right)$. 

The pathway amplitudes obviously do not scale asymptotically with the number of molecules $N$, they are on the order of $\mathcal{O}(1)$. However, their difference
\begin{equation}
    A-B = \frac{1}{2}\left(\frac{1}{2}+\sqrt{\frac{N}{2(2N-1)}}\right)^2 - \frac{1}{2}
\end{equation}
is on the order of $\mathcal{O}\left(N^{-1}\right)$. While the difference of the pathway amplitudes determines the asymptotic scaling of the trivial part of the approximated line shape function, the energy offset determines the scaling of the derivative-like part. Therefore, both contributions are on the order of $\mathcal{O}\left(N^{-1}\right)$. 

\bibliography{Disordered_lineshapes.bib}

\end{document}